\documentclass[12pt]{iopart}
\usepackage{graphicx}

\begin{document}

\title{Contagion spreading on complex networks with local dynamics}

\author{Pouya Manshour and Afshin Montakhab}

\address{Department of Physics, College of Sciences, Shiraz University, Shiraz 71454, Iran}
\ead{montakhab@shirazu.ac.ir}
\begin{abstract}
Typically, contagion strength is modeled by a transmission rate $\lambda$, whereby all nodes in a network are treated uniformly in a mean-field approximation. However, local agents react differently to the same contagion based on their local characteristics. Following our recent work [EPL \textbf{99}, 58002 (2012)], we investigate contagion spreading models with local dynamics on complex networks.  We therefore quantify contagions by their quality, $0 \leq \alpha \leq 1$, and follow their spreading as their transmission condition (fitness) is evaluated by local agents. We choose various deterministic local rules. Initial spreading with exponential quality-dependent time scales is followed by a stationary state with a prevalence depending on the quality of the contagion. We also observe various interesting phenomena, for example, high prevalence without the participation of the hubs. This is in sharp contrast with the general belief that hubs play a central role in a typical spreading process.  We further study the role of network topology in various models and find that as long as small-world effect exists, the underlying topology does not contribute to the final stationary state but only affects the initial spreading velocity.
\end{abstract}


\maketitle

\section{Introduction}
\label{intro}

Complex network theory has attracted much attention in recent years \cite{newmanbook,cohenbook,barabasibook,barrat2008}. In particular, dynamical processes on such topologies have been intensively studied and have helped understand many important processes in physics, biology, neuroscience, communications, epidemiology among others \cite{barrat2008,Castel2009,newman2002,murray1993,daley1999,hethcote2000,pastor2007,Dezso2002,Smilkov2012,Rohden2012,anderson1992}. Spreading of what is often called a contagion, e.g. virus or information, is one of the key areas of study with many important applications. Spreading, defined as transmission of a contagion from a node to its neighbors via links, is typically modeled as compartmentalization of individual nodes into a few categories S (susceptible), I (infected), R (recovered/removed) \cite{anderson1992,Grenfell1992,Garnett1996,keeling2008,renshaw1991,Dietz1967,kermack,bailey}. Most authors study such dynamics within a (heterogeneous) mean-field approximation where a global transition probability, $\lambda$, is assumed \cite{anderson1992,kermack,bailey,Dorogovtsev2008,Moreno2002,pastor2001,pastor2001-1,Gomez2011}.
It is found that even for small $\lambda$, a large part of the network is infected. This high prevalence occurs because of the key role played by highly connected nodes referred to as ``hubs'' \cite{Hethcote1984,barth2004,Keeling2005}. However, such  high prevalence for arbitrary $\lambda$ is in sharp contrast with real-world observations where most contagion spreading show low prevalence \cite{Kephart1993,Bai2007,White1998}.

We believe such inconsistencies can be traced back to the mean-field approach where the role of individual nodes and their interaction with various contagions is approximated in a uniform way (some exceptions include \cite{Olinky2004,Yang2008}). In other words, different nodes react differently to the same contagion \cite{Funk2010}. The infection by a virus depends strongly on the ``strength'' of the virus as well as the ``protection'' available at the receiving node. On the other hand, individuals typically evaluate the quality of the information according to their own values before deciding to pass them on. From a theoretical point of view, this requires construction of models of contagion dynamics which take into account the content of the contagion as well as the role of the individual agents receiving/evaluating/transmitting contagion.

We have recently \cite{Montakhab2012} proposed such a model where contagions are assigned strength or quality, $\alpha$, and their transmission only occurs after the local agents evaluate the appropriateness or quality of such contagions based on their local characteristics. We found low prevalence under generic conditions along with quasi-stationarity and power-law behavior in agent activities much in line with recent empirical results \cite{Lerman2010,Galuba2010}. In our previous work we considered a local rule which was inherently probabilistic. Stochastic elements of local dynamics were crucial in leading to quasi-stationarity and power-law behavior. Here, we propose to study \emph{deterministic} local rules. In particular, we study a deterministic version of our previous model as well as various other deterministic rules which characterize different class of spreading phenomena. The underlying topology is a key factor in various spreading phenomena. We therefore study the role played by various topologies in our model.
We find that the spreading process is characterized by an initial phase where contagions spread exponentially fast with time scales which depend on the quality of the contagion being spread. After the initial exponential phase, a quality-dependent fraction of the population is infected. This general behavior is shown to hold on various (complex) network structures regardless of topology, but depending crucially on the distribution of local quantities.

\section{Fitness-based local dynamics}
\label{sec:1}

We have recently introduced \cite{Montakhab2012} a model of contagion spreading where fitness criterion (transmission condition) was a probabilistic function of the incoming contagion quality. This was achieved by defining a Gaussian probability of transmission where the width set the scale for deviation of the contagion quality from the local variable. Here, we propose to study deterministic variations of such a model. But, we first briefly recall the original model introduced in \cite{Montakhab2012}. A local quantity $x(i)$ is introduced for each node $i$ indicating the quality of the node. Also, to quantify the quality of the contagion being spread, a parameter $\alpha$ is introduced, where $0\leq\alpha\leq1$. Each node interacts with the incoming contagion based on its fitness, i.e. how well $\alpha$ and $x(i)$ match. The interaction between the individuals and the incoming contagions is defined as follows:
At each time step, each node $i$ which receives the contagion $\alpha$, can accept or reject the contagion based on its observed fitness. If the contagion is accepted, the individual keeps that contagion forever and passes it on to all its $k_i$ neighbors. We initially assume that the quality of each individual is directly proportional to the number of its neighbors $k_i$,
\begin{equation}\label{Eq.1}
    x(i)=\frac{k_i}{k_{max}}
\end{equation}
where $k_{max}=max(k_i)$, making $0\leq x(i)\leq 1$ for any given network. Subsequently, we will consider quality distribution independent of network topology. We propose to consider three different local deterministic rules for acceptance (i.e. transmission) of contagions defined as follows:

I) Width-dependent condition (WDC): if $|\alpha-x(i)|\leq\Delta$, then accept the contagion otherwise deny it.

II) Threshold condition type 1 (TC1): if $\alpha\geq x(i)$, then accept the contagion otherwise deny it.

III) Threshold condition type 2 (TC2): if $\alpha\leq x(i)$, then accept the contagion otherwise deny it.

WDC is relevant when the local agent judges the quality of the incoming contagion and transmits it only if it is within a well-defined ($\Delta$) range of the local value. This local rule seems more applicable to information transmission as agents judge the quality  of the incoming information based on their own personal taste and transmit only if it matches their local values. This is a deterministic version of the local probabilistic rule we have considered previously \cite{Montakhab2012}. The threshold rules (TC1, TC2), on the other hand, may have more relevance in epidemiology as the strength of a certain contagion must reach a local threshold for immunity before infection and subsequent transmission occurs. For example, TC1 says that the virus strength must overcome the local threshold for immunization while TC2 indicates transmission ``under the (local) radar''. We propose to study such dynamics on both Erd\"{o}s-R\'{e}nyi (ER) \cite{Erdos1960} as well as scale-free (SF) \cite{Barabasi1999} networks. Thus a randomly selected node is seeded with a contagion of quality $\alpha$ and the subsequent dynamics is monitored.

\section{Analytical results}
\label{sec:2}
In this section we present some simple analytical results for the short time behavior of our model. Taking into account the heterogeneity introduced above on a network with connectivity distribution $p(k)$ and average connectivity $\left\langle k\right\rangle =\Sigma_{k}kp(k)$, one can write evolution equations for densities of informed or infected, $I_k (t)$, and uninformed or uninfected nodes, $U_k (t)$, leading to an \emph{early time} spreading, characterized by the average informed density $I(t)=\Sigma_kI_k(t)p(k)$ \cite{barth2004,Montakhab2012}:
\begin{equation}\label{Eq.2}
I(t)=I_0[1+\tau \left\langle kf(k,\alpha) \right\rangle (\frac{\left\langle k \right\rangle-1}{\left\langle k \right\rangle})(e^{t/\tau}-1)]
\end{equation}
where
\begin{equation}\label{Eq.3}
\tau=\frac{\left\langle k \right\rangle}{\left\langle k^2f(k,\alpha) \right\rangle-\left\langle kf(k,\alpha) \right\rangle}
\end{equation}
where $I_0\ll1$ is the initial informed density.

On the other hand, the local probability of acceptance depends on the local dynamics in the following way: \\

I) WDC:
\begin{equation}\label{Eq.4}
f(k,\alpha)=H((k/k_{max}+\Delta)-\alpha)+H(\alpha-(k/k_{max}-\Delta))-1
\end{equation}

II) TC1:
\begin{equation}\label{Eq.5}
f(k,\alpha)=H(\alpha-k/k_{max})
\end{equation}

III) TC2:
\begin{equation}\label{Eq.6}
f(k,\alpha)=H(k/k_{max}-\alpha)
\end{equation}
where $H(z)$ is the $Heaviside$ step function, i.e. $H(z)=1$ for $z\geq1$ and $H(z)=0$ otherwise.

The exponential growth in early times (equation \ref{Eq.2}) is typical of SI models. Equation \ref{Eq.3}, however, is an interesting result. It shows that the growth time scale in our model is dependent on local dynamics as well as the topology of the network. Perhaps more importantly, it shows that not only different classes of local dynamics, i.e. $f(k,\alpha)$, show different behavior, but such a behavior within each class is dependent on the value of $\alpha$. In the next section, we will numerically check the validity of equation \ref{Eq.3} by directly simulating various models. We will also show how different model classes have distinctly different spreading states depending on $\alpha$, the quality of the contagion being spread. We also note that our model is a generalization of the standard SI model for contagion spreading, with a general global transmission probability $f(k,\alpha)=\lambda$, reducing equation \ref{Eq.3} to the well-known previous results \cite{barth2004,Montakhab2012}. Here, the transmission probability has an explicit dependence on the strength of contagion as well as the local agents involved in the transmission process.

\section{Numerical results}
\label{sec:3}
We have performed numerical simulations of our model on two types of networks, SF with $p(k)\propto k^{-\gamma}$ ($\gamma=2.5$ or $\gamma=3$) and an ER network with a \emph{Poisson} degree distribution. The network realizations used for the numerical simulations were constructed using the method introduced in \cite{Catanzaro2005} in order to assure that no degree-degree correlations are present in any of the networks generated. We first pick a randomly chosen node and seed it with contagion $\alpha$ and then monitor various quantities of interest as spreading takes place. We average over at least $1000$ different starting configurations, and consider at least $10$ different realizations for each given network. In what follows it is useful to define $\left\langle x\right\rangle=\left\langle k\right\rangle/k_{max}$ as the average quality of a given network.
\\
\begin{figure}
  \includegraphics{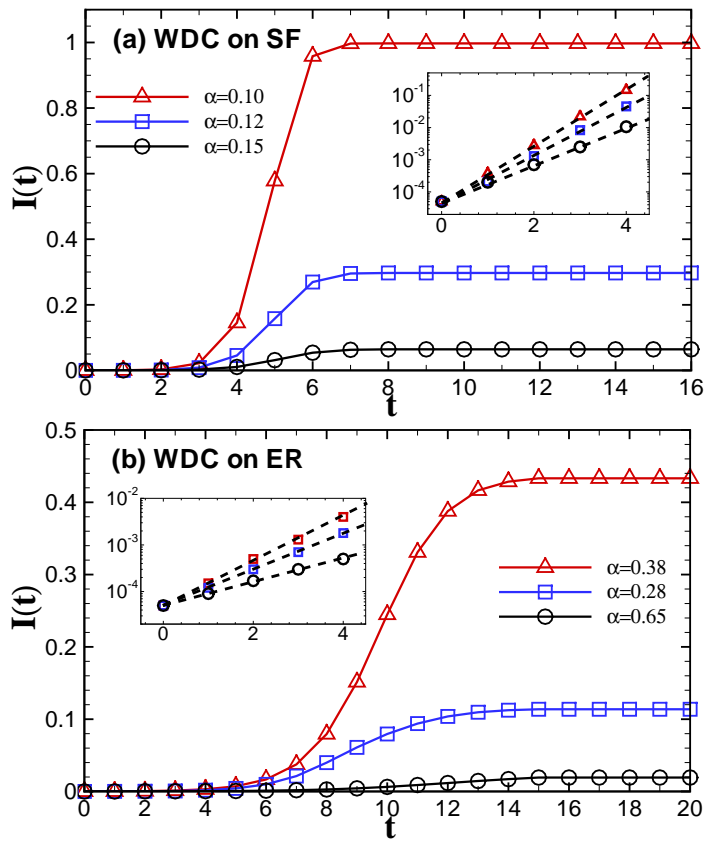}
\caption{Averaged density of informed nodes versus time for WDC ($\Delta=0.1$) on (a) a SF network of $\gamma=3$ and (b) an ER network. The networks have the same size of $N=20000$ and average degree $\left\langle k\right\rangle=6$. The insets show the exponential growth during the first few steps.}
\label{Fig1}
\end{figure}

\begin{figure}
  \includegraphics{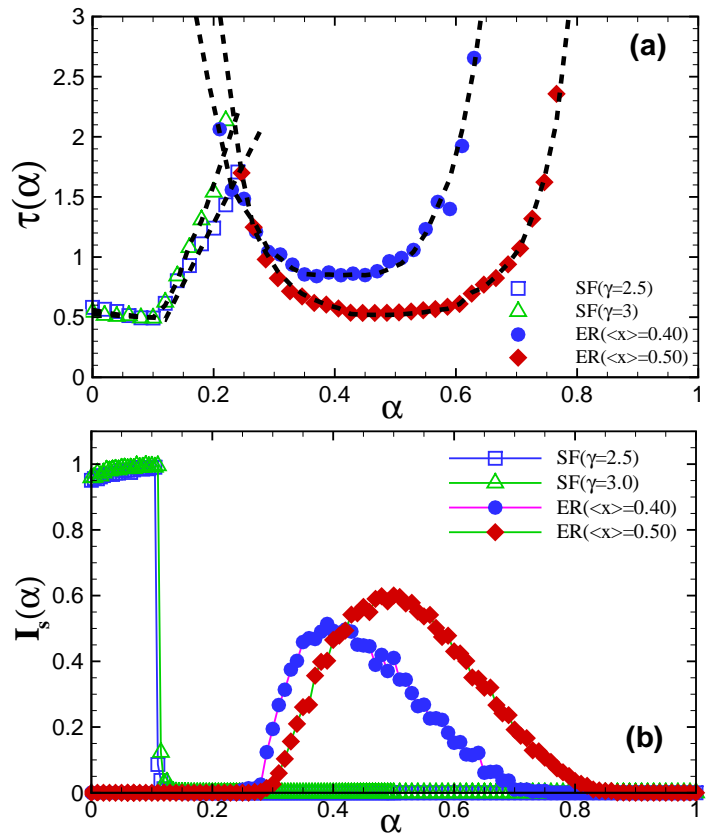}
\caption{(a) Measured time scale $\tau$ vs $\alpha$ as obtained from exponential fitting and (b) the stationary state of the density of informed nodes versus contagion $\alpha$ for WDC ($\Delta=0.1$) on two SF networks of $\gamma=2.5$ (squares) and $3$ (triangles) with mean degree $\left\langle k\right\rangle=6$ and two ER networks with mean degrees $\left\langle k\right\rangle=6$ (bullets) and $\left\langle k\right\rangle=12$ (diamonds). Dashed lines in (a) indicate the theoretical prediction (equation \ref{Eq.3}). The system size is $N=20000$ for all networks.}
\label{Fig2}
\end{figure}

\begin{figure}
\includegraphics{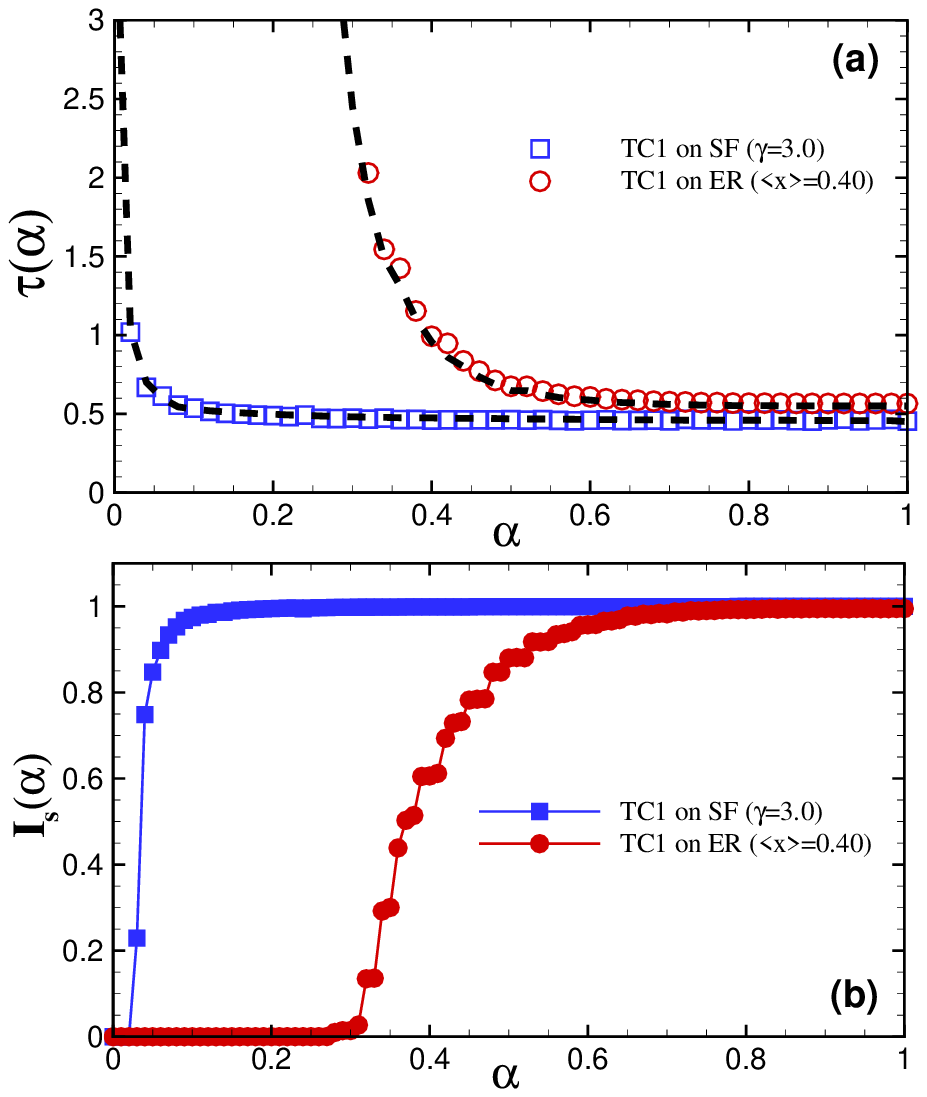}
\caption{(a) Measured time scale $\tau$ vs $\alpha$ as obtained from exponential fitting and (b) the stationary state of the density of informed nodes versus contagion $\alpha$ for TC1 on a SF network of $\gamma=3$ (squares) and an ER graph (circles) both with the same mean degree $\left\langle k\right\rangle=6$. Dashed lines in (a) indicate the theoretical prediction (equation \ref{Eq.3}). The system size is $N=20000$ for both networks.}
\label{Fig3}
\end{figure}

\begin{figure}
\includegraphics{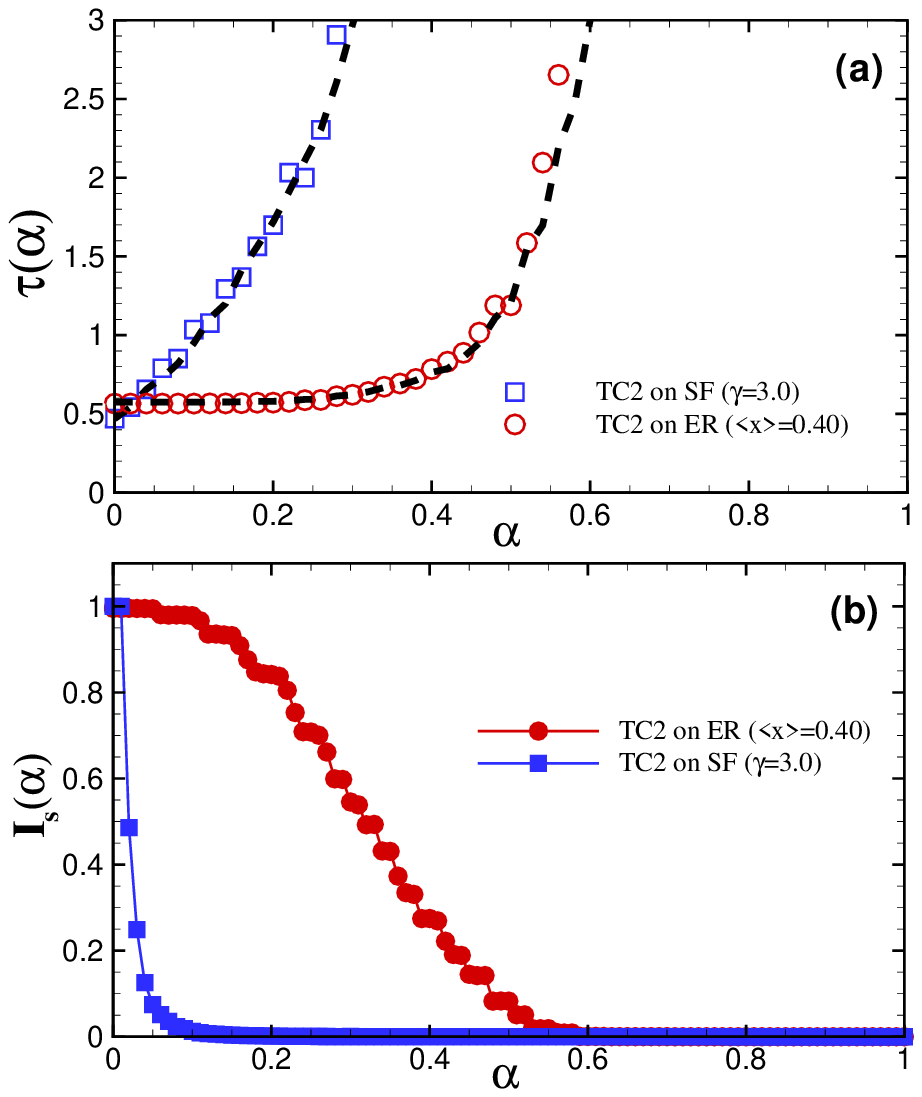}
\caption{(a) Measured time scale $\tau$ vs $\alpha$ as obtained from exponential fitting and (b) the stationary state of the density of informed nodes versus contagion $\alpha$ for TC2 on a SF network of $\gamma=3$ (squares) and an ER graph (circles) both with the same mean degree $\left\langle k\right\rangle=6$. Dashed lines in (a) indicate the theoretical prediction (equation \ref{Eq.3}). The system size is $N=20000$ for both networks.}
\label{Fig4}
\end{figure}

\subsection{WDC} Figure \ref{Fig1} shows the time evolution of average density of informed nodes $I(t)$ for WDC rule with $\Delta=0.1$ for three different values of contagion $\alpha$ on a SF network of exponent $\gamma=3$ (figure \ref{Fig1}(a)) and an ER network (figure \ref{Fig1}(b)) with the same average degree $\left\langle k\right\rangle =6$ and network size $N=20000$. Insets show a log-linear plot of the early time behavior of $I(t)$ indicating an exponential growth with $\alpha$-dependent time scales. As can be seen from figure \ref{Fig1}, the process of spreading is characterized by an initially fast exponential growth and eventual approach to a stationary state. In the final stationary state a certain fraction of the nodes are informed which clearly depend on the value of $\alpha$ for a given $\Delta$.
figure \ref{Fig2}(a) shows the growth time-scale $\tau$ obtained from exponential fitting of the numerical results along with the theoretical prediction equation \ref{Eq.3}, versus $\alpha$ for two types of network. We note that the numerical results recover the analytical calculation with good accuracy. In figure \ref{Fig2}(b) we report stationary state of the density of informed nodes, $I_s=I(t\rightarrow\infty)$, versus contagion $\alpha$ for WDC model. We see that the fastest spreading (smallest $\tau$) coincides with largest spreading (largest $I$). For such a model, the most efficient spreading occurs when the quality of contagion matches the average quality of the network $\alpha=\left\langle x\right\rangle$. Significant spreading also occurs for a width of $\Delta$ about such maximum, i.e., $\alpha=\left\langle x\right\rangle\pm\Delta$. However, for SF network $\left\langle x\right\rangle=\left\langle k\right\rangle/k_{max}\rightarrow 0$ as $k_{max}$ is arbitrary large for large networks, while for ER network a finite $\left\langle x\right\rangle$ typically exists. As is seen in figure \ref{Fig2}(b), the cutoff is sharp on SF network while it is gradual on ER network which can be understood in terms of their connectivity distribution $p(k)\propto p(x)$. In both types of networks the final size of informed nodes is proportional to the number of nodes whose quality matches the quality of the contagion $\alpha$ within $\Delta$. Therefore, low quality contagions spread extremely efficiently on SF networks ($I_s\approx1$) with a sharp cut-off to a non-prevalent state at $\Delta$, while efficient spreading on ER network must be accompanied by targeting the contagion to fall around the average quality of the network with relatively lower prevalent state (e.g. $I_s \approx 0.6$). Our results may have important implications for marketing/advertising strategies. For example, low quality information, e.g. tabloids, spread well in a heterogenous society while they remain localized in a homogenous one.  Furthermore, we note that quasi-stationarity and power law behavior in agent activity seen in the stochastic model \cite{Montakhab2012} is no longer observed in the deterministic model as such properties crucially depend on the stochastic nature of the local agent.

\subsection{TC1} This condition allows for transmission only if the contagion $\alpha$ overcomes the local threshold, i.e., $\alpha\geq x(i)$. An example is when a computer virus must overcome the protective softwares installed locally. Figure \ref{Fig3} shows the growth time-scale $\tau$ (figure \ref{Fig3}(a)) along with the theoretical prediction equation \ref{Eq.3}, and the stationary state of the density of informed nodes versus contagion $\alpha$ (figure \ref{Fig3}(b)) for TC1 rule on both networks. Since transmission occurs only when $\alpha\geq x(i)$ locally, one expects no significant transmission for the limit $\alpha\rightarrow0$, a point well characterized by the divergence of $\tau$ as $\alpha\rightarrow0$ (figure \ref{Fig3}(a)) and its correspondence to no prevalence ($I_s\approx0$) in figure \ref{Fig3}(b). Again such transition is sharper on the SF network and occurs around $\left\langle x\right\rangle$ which remains finite on ER network, $\left\langle x\right\rangle=0.40$ in our case, but goes to zero for large SF networks implying complete prevalence for all $\alpha$ (except $\alpha=0$). The high prevalence ($I_s\approx1$) observed in TC1 model is typical of epidemic models frequently studied in the literatures, in particular on SF networks. However, we note that the actual dynamical process is distinctly different from the spreading process for, say, the SI model on SF network. For example, here the high prevalence state is reached without ever infecting the hubs, while in the standard models the hubs are the ``super-spreaders" \cite{Hethcote1984,barth2004,Keeling2005}. This is an important distinction as it is commonly believed that hubs play a crucial role in high-prevalence spreading processes, and thus their ``removal" is an effective immunization strategy.  Our results show this not to be the case.

\subsection{TC2} This condition is the opposite of TC1. It allows for transmissions unless $\alpha$ exceeds the local quality $x(i)$. This is relevant to ``moving under a radar'' where local agents stop transmissions once they detect the contagion according to their predefined local detector, $x(i)$. The results for such a model are shown in figure \ref{Fig4} and, as expected, display the opposite behavior of TC1 in figure \ref{Fig3}. We note that ``moving under the radar'' is almost impossible on SF networks while it is quite effective on an ER network as long as $\alpha\leq \left\langle x\right\rangle$.

\section{Topology and small-world effect}
\label{sec:4}
Our results so far have indicated that the characteristics of our model dynamics strongly depends on network topology as plots of $\tau(\alpha)$ and $I_s(\alpha)$ show  distinctly different behavior on homogenous and heterogeneous topologies. However, since we have assumed $p(x)\propto p(k)$ so far, we have tied the distribution of quality to our topology. In this section, we de-couple such distribution in order to better address their separate effects on the spreading processes studied here. One can easily see that the important factor determining the final stationary form $I_s(\alpha)$ is the distribution of the local variables $x(i)$, as the previous profiles (figure \ref{Fig2}(b), figure \ref{Fig3}(b) and figure \ref{Fig4}(b)) are effectively the percentage of the
nodes whose local variable meets the transmission condition.  It seems like that the actual role of the underlying topology should not be important as long as small-world effect is present.  The small-world effect guarantees that all nodes whose local variable are ``fit" are exposed and thus infected. On the other hand, one would expect that the role of different topologies would have a direct effect on the actual dynamical process leading to the final stationary state and would therefore directly effect short-time dynamics, i.e. $\tau(\alpha)$.

\begin{figure}
\includegraphics{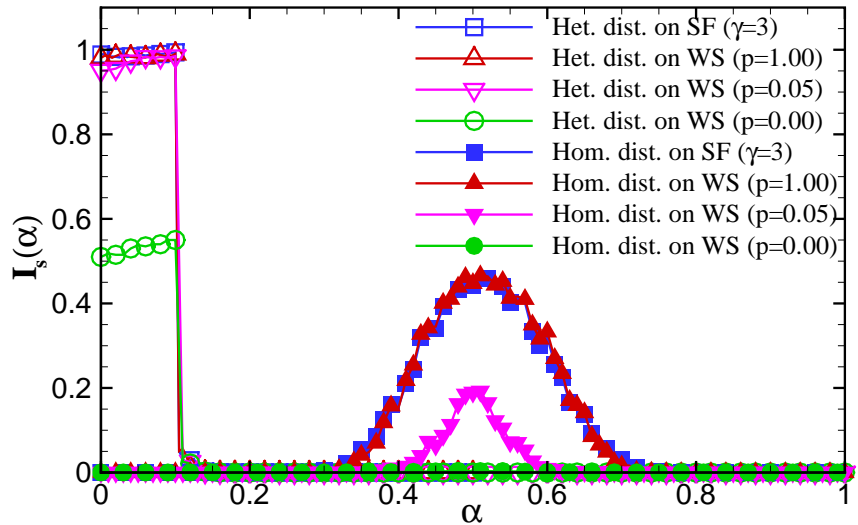}
\caption{The stationary states of the density of informed nodes versus contagion $\alpha$ for WDC ($\Delta=0.1$) for heterogeneous distribution (open symbols) and homogeneous distribution (filled symbols) on a SF network of $\gamma=3$ (squares) and three WS networks with $p=1.0$ (deltas), $p=0.05$ (gradients) and $p=0.0$ (circles), all with the same average degree $\left\langle k\right\rangle=4$. The network size is $N=5000$.}
\label{Fig5}
\end{figure}

\begin{figure}
\resizebox{0.65\textwidth}{!}{%
\includegraphics{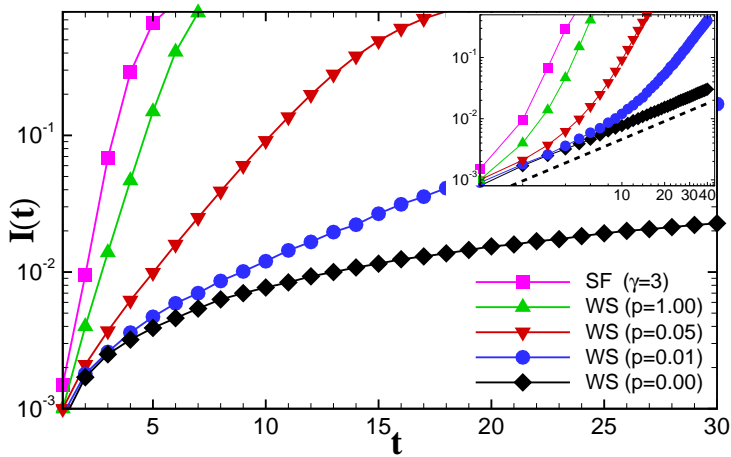}
}
\caption{Log-linear plots of the early time behavior of the density of informed nodes for WDC ($\Delta=0.1$) for heterogeneous distribution on a SF network of $\gamma=3$ (squares) and four WS networks with $p=1.0$ (deltas), $p=0.05$ (gradients), $p=0.01$ (circles) and $p=0.0$ (diamonds). Inset is the log-log plots of the figure. Dashed line in the inset is a line with slope $1$. The networks parameter are the same as in figure \ref{Fig5}.}
\label{Fig6}
\end{figure}

In order to check the above arguments, we have produced a homogeneous (Poissonian) as well as a heterogeneous (scale-free) distribution of $x(i)'s$ and have studied their random distribution on various network topologies.
We do this by considering a scale-free network as well as a Watts-Strogatz (WS) \cite{Watts1998,Watts1999} small-world network which is characterized by a rewiring probability, $p$, which allows us to extrapolate from a regular network with no small-world effect ($p=0$) to a complete random (ER) network with $p=1.0$. We show our simulation results for the WDC rule in figure \ref{Fig5}. We note that for the heterogenous distribution in the presence of a network with small-world effect the profiles are identical regardless of network topology. For the homogenous distribution, however, for both ER and SF topologies we obtain identical profiles while the profile for the $p=0.05$ shows a similar but yet smaller prevalence. The case of regular network, $p=0.0$, is capable of distinguishing between the homogeneous and heterogeneous distributions. Heterogeneity allows for significant spreading in the appropriate regime (small $\alpha$) even in the absence of small-world effect, while homogeneity leads to localized activities and lack of prevalence on a regular network regardless of the value of $\alpha$. The significant spreading observed on a regular network with heterogeneous distribution is interesting and is related to wave-like spreading previously seen in other models on regular lattice \cite{Grenfell2001,Mollison1977,Rhodes1997}. To show such dynamical process better, we have plotted the initial spreading process for the heterogeneous distribution on various network structures in figure \ref{Fig6}. We see that for $p=0.0$ the spreading follows a power law (linear) growth indicative of wave-like spreading, while in the presence of small-world effect follows an exponential growth with scales depending on the underlying topologies. The linear profile has to do with the one-dimensional nature of the WS network with $p=0.0$. Therefore, while the initial phase of spreading is effected by the underlying topology, the final stationary state is relatively independent of the underlying topology given small-world effect, more so for the heterogeneous population than a homogenous one.

\section{Conclusion}
\label{sec:5}
In this work we have studied contagion spreading on complex networks with local deterministic dynamics. Our local dynamics is based on evaluation of the fitness of the contagion by the local agent in allowing its subsequent transmission. We previously studied a stochastic local rule where we observed quasi-stationarity and power-law behavior in agent activity. Here, we study various deterministic rules where such a behavior is no longer expected. We have considered a WDC rule which is a deterministic version of our previous model and also study threshold conditions (TC1 and TC2) more relevant to epidemiology. Regardless of dynamical rules we observe exponential growth with contagion ($\alpha$) dependent time scales leading to a stationary  state where medium or low prevalence is observed for a wide range of parameters, thus being more consistent with empirical observations. In fact, the final prevalence, $I_s(\alpha)$, is shown to be the relative number of agents who meet the fitness criterion associated with the given $\alpha$. This behavior was shown to be independent of network topology as long as significant small-world effect exists, while topology was shown to have an effect on early time dynamics. Finally, we mention that our WDC model which is more relevant to information spreading guaranteed high prevalence for low quality contagions in a heterogeneous network while for homogeneous networks targeting the contagion was required to achieve significant prevalence. This result seems particularly important for marketing/advertising strategies. On the other hand, for our threshold condition TC1 relevant to virus spreading, we observed high prevalence for a wide range of $\alpha$ (particularly in a heterogeneous distribution) without the participation of the hubs. This result is in contrast with previous studies and may be important for immunization strategies.

\section*{Acknowledgments}
Support from Shiraz University Research Council is kindly acknowledged.

\end{document}